\documentclass[english,aps,preprint,nofootinbib,showpacs,showkeys,superscriptaddress]{revtex4}
\usepackage[T1]{fontenc}
\usepackage[latin9]{inputenc}
\usepackage{textcomp}
\usepackage{bm}
\usepackage{amsmath}
\usepackage{graphicx}
\usepackage{amssymb}

\makeatletter
\@ifundefined{textcolor}{}
{%
 \definecolor{BLACK}{gray}{0}
 \definecolor{WHITE}{gray}{1}
 \definecolor{RED}{rgb}{1,0,0}
 \definecolor{GREEN}{rgb}{0,1,0}
 \definecolor{BLUE}{rgb}{0,0,1}
 \definecolor{CYAN}{cmyk}{1,0,0,0}
 \definecolor{MAGENTA}{cmyk}{0,1,0,0}
 \definecolor{YELLOW}{cmyk}{0,0,1,0}
 }
\newenvironment{lyxlist}[1]
{\begin{list}{}
{\settowidth{\labelwidth}{#1}
 \setlength{\leftmargin}{\labelwidth}
 \addtolength{\leftmargin}{\labelsep}
 }}
{\end{list}}

\usepackage{bbm}
\usepackage{bm}
\usepackage{hyperref}
\usepackage{slashed}

\makeatother

\usepackage{babel}

\begin{document}
\global\long\def\half{\frac{1}{2}}
\global\long\def\Imag#1{\mathrm{Im}\left\{  #1 \right\}  }
\global\long\def\Real#1{\mathrm{Re}\left\{  #1 \right\}  }
\global\long\def\db{\!\not\!\! D\,}
\global\long\def\abs#1{\left|#1\right| }
\global\long\def\then{\quad\Rightarrow\quad}
\global\long\def\lcal{\mathcal{L}}
\global\long\def\mcal{\mathcal{M}}
\global\long\def\sigbf{\bm{\sigma}}
\global\long\def\inv#1{\frac{1}{#1}}
\global\long\def\nurp{{\nup_{R}}}
\global\long\def\nul{{\nu_{L}}}
\global\long\def\nulp{{\nup_{L}}}
\global\long\def\sla#1{\rlap{$#1$}/}
\global\long\def\er{e_{R}}
\global\long\def\el{e_{L}}
\global\long\def\nup{ \nu^{\prime} }
\global\long\def\nur{{\nu_{R}}}
\global\long\def\mnp{M_{\mathrm{NP}}}
\global\long\def\lnp{\Lambda_{\mathrm{NP}}}
\global\long\def\cw{c_{{\rm W}}}
\global\long\def\sw{s_{{\rm W}}}
\global\long\def\mix{\eta}
\global\long\def\notcp{\slashed{\mathrm{CP}}}

\preprint{FTUV-09-0421, IFIC-09-15, UCRHEP-T466}

\title{Right-handed neutrino magnetic moments}

\author{Alberto Aparici}

\affiliation{Departament de Física Teòrica, Universitat de València\\
 and IFIC, Universitat de València-CSIC\\
Dr. Moliner 50, E-46100 Burjassot (València), Spain}

\author{Kyungwook Kim }

\affiliation{Department of Physics and Astronomy, University of California, Riverside
CA 92521-0413, USA}

\author{Arcadi Santamaria}

\affiliation{Departament de Física Teòrica, Universitat de València\\
 and IFIC, Universitat de València-CSIC\\
Dr. Moliner 50, E-46100 Burjassot (València), Spain}

\author{José Wudka}

\affiliation{Department of Physics and Astronomy, University of California, Riverside
CA 92521-0413, USA}
\begin{abstract}
We discuss the phenomenology of the most general effective Lagrangian,
up to operators of dimension 5, build with standard model fields and
interactions including right-handed neutrinos. In particular we find
there is a dimension 5 electroweak moment operator of right-handed
neutrinos, not discussed previously in the literature, which could
have interesting phenomenological consequences. 
\end{abstract}



\maketitle

\section{Introduction\label{sec:Introduction}}

Since the first hints on neutrino masses~\cite{Davis:1968cp,Hirata:1988uy},
the physics of neutrinos is coming of age with a significant amount
of new and increasingly precise data and a variety of new experiments.
Though a significant number of parameters in the neutrino sector have
been recently measured~\cite{Cleveland:1998nv,Hampel:1998xg,Altmann:2000ft,Ahmad:2002jz,Eguchi:2002dm,Fukuda:1998mi,Allison:1999ms,Aliu:2004sq}
(for a recent global fit see~\cite{Schwetz:2008er} and for a recent
review see~\cite{GonzalezGarcia:2007ib}), many questions remain.
In particular it is not known whether neutrinos are (dominantly) Dirac
or Majorana fermions, what is their absolute mass scale and whether
they have the electromagnetic properties predicted by the Standard
Model. In this paper we will concentrate mostly on the latter issue
(for a very recent review see~\cite{Giunti:2008ve}) .

Given that our knowledge of neutrino interactions is limited, it is
sensible to study neutrino properties using a framework that includes
possible non-SM interactions in a systematic way. This is most easily
done using an effective Lagrangian. The application of this formalism
to the neutrino system exhibits novel complications since the complete
set of low-energy degrees of freedom is not definitively known. For
example, the appropriate description of the light neutrino masses
may require the introduction of new relatively light ($\lesssim$TeV)
degrees of freedom%
\footnote{For instance, to explain baryon asymmetry and dark matter in the universe
one may need light right-handed neutrinos~\cite{Laine:2008pg}.%
}, which might be convenient to include in the low-energy theory, and
the approach must be sufficiently general to allow for this possibility.

The effective Lagrangian approach is reliable only at energies significantly
below the scale of new physics~\cite{Coleman:1969sm,Weinberg:1978kz,Weinberg:1980wa,Polchinski:1983gv,Georgi:1994qn}
that will be denoted by $\mnp$. In addition we will assume that the
underlying physics is decoupling~\cite{Appelquist:1974tg}, so that
the effective theory can be expanded in powers of $1/\mnp$. The use
of effective theories in neutrino physics is far form new~\cite{Weinberg:1979sa,Weldon:1980gi}
(for recent applications see, for instance, \cite{Bilenky:1993bt,Bilenky:1994ma,Grossman:1995wx,Bergmann:1999rz,Babu:2001ex,Berezhiani:2001rs,Oliver:2001eg,Broncano:2002rw,Davidson:2003ha,Prezeau:2003xn,Barranco:2005ps,Davidson:2005cs,delAguila:2007ap});
despite this we find that when right-handed neutrinos are included
in the low-energy theory, not all the interactions allowed by gauge
invariance have been adequately studied in the literature~\cite{delAguila:2008ir}.

The first-order corrections (in powers of $1/\mnp$) to the SM interactions
correspond to dimension 5 operators, which in our case fall into three
classes: those contributing to the Majorana mass matrices for the
left and right-handed neutrinos, and those describing a magnetic moment
coupling for the right-handed neutrinos; it is this last term that
has been largely ignored.

In the following we will investigate several properties and consequences
of this new electroweak interaction and discuss its origin, experimental
constraints and possible effects both in collider experiments and
in various areas of astrophysics and cosmology.

\section{Dimension 5 effective Lagrangian\label{sec:leff}}

When considering the low energy effects of a (hypothesized) heavy
physics that is not directly probed, it is convenient to parametrize
all new physics effects using a series of effective vertices involving
only light fields~\cite{Weinberg:1980wa,Georgi:1994qn,Wudka:1994ny}.
These vertices are constrained only by the gauge invariance of the
light theory~\cite{Veltman:1980mj}. Assuming that the physics underlying
the Standard Model (SM) is decoupling, the heavy-physics corrections
to the SM processes will be suppressed by powers of the heavy scale~%
\footnote{Though there are corrections that grow with $\mnp$ these can always
be absorbed in the renormalization of the SM parameters. Even if formally
unobservable, these contributions are of interest when the naturality
of the theory is studied.%
} $\mnp$.

Concerning the light degrees of freedom, we will assume these consist
of all the SM excitations together with 3 right-handed neutrinos $\nurp$,
assumed to be gauge singlets (the prime indicates that these are not
mass eigenstates). Should the scale of the $\nurp$ be $\gtrsim\mnp$,
these excitations will disappear from the low-energy theory; the effective
theory in this case is obtained from the expressions below by simply
erasing all Lagrangian terms containing the $\nurp$.

The most general form of the effective Lagrangian including up to
dimension 5 terms is~

\begin{eqnarray}
\lcal & = & \lcal_{\mathrm{SM}}+\lcal_{\mathrm{\nu_{R}}}+\lcal_{\mathrm{5}}+\cdots\\
\lcal_{\mathrm{SM}} & = & i\overline{\ell}\db\ell+i\overline{\er}\db\er-(\overline{\ell}Y_{e}e_{L}\,\phi+\mathrm{h.c.})+\cdots\label{eq:LagrangianSM}\\
\lcal_{\nur} & = & i\overline{\nu_{R}^{\prime}}\sla{\partial}\,\,\nu_{R}^{\prime}-\left(\half\overline{\nu_{R}^{\prime c}}M\nurp\,+\mathrm{h.c.}\right)-\left(\overline{\ell}Y_{\nu}\nu_{R}^{\prime}\,\tilde{\phi}+\mathrm{h.c.}\right)\label{eq:LagrangianSMnur}\\
\lcal_{\mathrm{5}} & = & \overline{\nu_{R}^{\prime c}}\zeta\sigma^{\mu\nu}\nu_{R}^{\prime}B_{\mu\nu}+\left(\overline{\tilde{\ell}}\phi\right)\chi\left(\left.\tilde{\phi}\right.^{\dagger}\ell\right)-\left(\phi^{\dagger}\phi\right)\overline{\nu_{R}^{\prime c}}\xi\nurp+\mathrm{h.c.}\label{eq:L5}\end{eqnarray}
where $\ell={\nu_{L}^{\prime} \choose e_{L}}$ denotes the left-handed
lepton isodoublet, $e_{R}$ and $\nu_{R}^{\prime}$ the corresponding
right-handed isosinglets, and $\phi$ the scalar isodoublet (family
and gauge indices will be suppressed when no confusion can arise);
we will assume three right-handed neutrino flavors. The charge-conjugate
fields are defined as $\er^{c}=C\bar{e}_{R}^{T},\:\nu_{R}^{\prime c}=C\bar{\nu}_{R}^{\prime T}$
and $\tilde{\ell}=\epsilon C\bar{\ell}^{\, T},\,\tilde{\phi}=\epsilon\phi^{*}$
where $\epsilon=i\sigma_{2}$ acts on the $SU(2)$ indices. The hypercharges
assignments are $\phi:\,1/2$, $\ell:-1/2$, $e_{R}:-1$, $\nu_{R}^{\prime}:\,0$.
The $SU(2)$ and $U(1)$ gauge fields are denoted by $W$ and $B$
respectively (gluon and quarks fields will not be needed in the situations
considered below). The Yukawa couplings $Y_{e}$ and $Y_{\nu}$ are
completely general $3\times3$ matrices in flavor space; $M$, $\chi$,
and $\xi$ are complex symmetric $3\times3$ matrices in flavor space
that generate the most general neutrino mass matrix, while $\zeta$
is a complex antisymmetric matrix proportional to the right-handed
neutrino electroweak moments. Without loss of generality, $Y_{e}$
and $M$ can be taken diagonal with positive and real elements.

The term involving $M$ is the usual right-handed neutrino Majorana
mass. The term involving $\chi$ was first described by Weinberg~\cite{Weinberg:1979sa}
and provides a Majorana mass for the left-handed neutrino fields plus
various lepton-number-violating neutrino-Higgs interactions; this
type of effective operator is the same that is obtained when considering
generic see-saw models. The term involving $\zeta$ has been mostly
ignored in the literature; it describes electroweak moment couplings
of the right-handed neutrinos. We will dedicate a significant part
of this paper to the study of some of the consequences this operator
might have on various collider, astrophysical an cosmological observables.
Note that Dirac-type neutrino magnetic moments (involving $\ell$
and $\nu_{R}^{\prime}$) are generated by operators of dimension $\ge6$,
while Majorana-type magnetic moments for left-handed neutrinos (involving
only the $\ell$) require operators of dimension $\ge7$. One can
easily see that these effects are subdominant when compared to those
produced by the term containing $\zeta$ in $\lcal_{5}$. In addition,
Majorana-type and Dirac-type magnetic moment operators contribute,
at the loop level, to neutrino masses~\cite{Davidson:2005cs,Bell:2005kz}
and, therefore, are strongly constrained. 

The couplings $\chi$, $\xi$, $\zeta$ have dimension of inverse
mass, which is associated with the scale of the heavy physics responsible
for the corresponding operator. Though we will refer to this scale
generically as $\mnp$ it must be kept in mind that different types
of new physics might be responsible for the various dimension 5 operators
and that the corresponding values of $\mnp$ might be very different.
One common characteristic of all these scales is that they should
all be much larger than the electroweak scale $v\sim0.25\,\mathrm{TeV}$,
by consistency of the approach being used. Below we discuss the possible
types of new physics that can generate these operators and the natural
size for the corresponding coefficients.

\subsection{Heavy-physics content of the effective vertices.}

As mentioned previously there are various kinds of heavy physics that
can generate $\lcal_{5}$ at low energies; we will briefly discuss
the various possibilities.

\subsubsection{$\nu_{L}$ Majorana mass term.}

Using appropriate Fierz transformations we can re-write the operator
containing $\chi$ as follows ($i$ and $j$ denote family indices):
\begin{equation}
\left(\overline{\tilde{\ell}_{i}}\phi\right)\left(\left.\tilde{\phi}\right.^{\dagger}\ell_{j}\right)=-\left(\overline{\tilde{\ell}_{i}}\sigbf\phi\right)\cdot\left(\left.\tilde{\phi}\right.^{\dagger}\sigbf\ell_{j}\right)=\half\left(\overline{\tilde{\ell}_{i}}\sigbf\ell_{j}\right)\cdot\left(\left.\tilde{\phi}\right.^{\dagger}\sigbf\phi\right)\,.\end{equation}
 It follows that this operator can be generated perturbatively at
tree level by the exchange of \textit{(i)} a scalar isotriplet of
hypercharge $1$, \textit{(ii)} a zero hypercharge fermion isotriplet,
or \textit{(iii)} a fermion isosinglet also of zero hypercharge (note
that these are the quantum numbers of the $\nu_{R}$, which are required
in many extensions of the SM). For weakly coupled heavy physics we
then expect

\begin{equation}
\chi\sim\lambda^{2}/\mnp,\label{eq:chi.est}\end{equation}
 where $\mnp$ denotes the mass of the corresponding heavy particle
and $\lambda$ the coupling constants of the heavy fermions to $\phi\ell$,
or of the heavy scalar to $\phi\phi$ and $\ell\ell$. When generated
by a scalar (fermion) isotriplet this interaction can realize the
type II (III) see-saw mechanism~\cite{Kobzarev:1980nk,Bilenky:1980cx,Cheng:1980qt,Schechter:1980gr,Ma:1998dx,Mohapatra:1991ng,Mohapatra:1980yp,Lazarides:1980nt,Magg:1980ut};
when generated by singlet neutrinos it realizes type I see-saw~\cite{Minkowski:1977sc,Ramond:1979py,GellMann:1980vs,Yanagida:1979as,Mohapatra:1979ia}.

\subsubsection{$\nu_{R}$ Majorana mass term.\label{sec:maj.mass} }

The operator \begin{equation}
\left(\phi^{\dagger}\phi\right)\overline{\nu_{iR}^{\prime c}}\nu_{jR}^{\prime}\end{equation}
can be generated at tree level by \textit{(i)} a scalar isosinglet
of vanishing hypercharge, or \textit{(ii)} a fermion isodoublet of
hypercharge $1/2$. We again expect \begin{equation}
\xi\sim\lambda^{2}/\mnp,\label{eq:xi.est}\end{equation}
 where $\mnp$ again denotes the mass of the heavy particles, and
$\lambda$ the coupling of the heavy fermion to $\phi\nu$ or the
heavy scalar to $\phi^{\dagger}\phi$ and $\nu\nu$ . Except for the
neutrino-Higgs interactions, the effects of this operator can be absorbed
into a redefinition of the Majorana mass $M$. The terms that do involve
the Higgs particle may open a new decay channel $H\to NN$ for the
Higgs boson (provided it is kinematically allowed). We will discuss
this possibility in section~\ref{sec:collider}.

\subsubsection{$\nu_{R}$ electroweak coupling.}

Finally, the operator \begin{equation}
\left(\overline{\nu_{iR}^{\prime c}}\sigma^{\mu\nu}\nu_{jR}^{\prime c}\right)B_{\mu\nu}\end{equation}
can be generated only at the one loop level by \textit{(i)} a scalar-fermion
pair $\{\omega,E\}$, with opposite (non-zero) hypercharges that have
couplings $\omega\overline{E}\nu_{R}^{\prime}$ and $\omega\overline{E}\nu_{R}^{\prime c}$,
or \textit{(ii)} a vector-fermion pair $\{W_{\mu}^{\prime},E\}$,
with opposite (non-zero) hypercharges that have couplings $W_{\mu}^{\prime}\overline{E}\gamma^{\mu}\nu_{R}^{\prime}$
and $W_{\mu}^{\prime}\overline{E}\gamma^{\mu}\nu_{R}^{\prime c}$.
Then \begin{equation}
\zeta\sim\frac{g^{\prime}y\lambda^{2}}{16\pi^{2}}\frac{m_{\mathrm{fermion}}}{\max(m_{\mathrm{fermion}}^{2},m_{\mathrm{boson}}^{2})}<\frac{g^{\prime}y\lambda^{2}}{16\pi^{2}m_{\mathrm{fermion}}}\,,\label{eq:seta.est}\end{equation}
where $\lambda$ denotes the coupling of the two heavy particles to
the $\nu_{R}^{\prime}$, and $y$ the hypercharge of the heavy boson
or fermion. A specific example is provided in appendix~\ref{sec:mod.calc}.

\vspace*{\bigskipamount}

We should mention that these coefficient estimates need not hold in
case the underlying physics is strongly coupled. In this case one
can obtain a natural estimate for the various coefficients using naive
dimensional analysis (NDA)~\cite{Manohar:1983md,Georgi:1992dw}.
The resulting values are \begin{equation}
\chi,\xi\sim\frac{16\pi^{2}}{\mnp};\qquad\zeta\sim\frac{1}{\mnp}\,,\label{eq:strong.estimate}\end{equation}
where $\mnp$ is, in this case, the scale of the strong interactions;
it is important to note that these estimates are based on the assumption
that $\ell,~\nu$ and $\phi$ participate in these strong interactions.
It is also worth noting that these estimates revert to the previous
ones (\ref{eq:chi.est}), (\ref{eq:xi.est}) and (\ref{eq:seta.est})
upon replacing $\mnp\to(4\pi)^{2}\mnp$.

In the following we will denote by $\lnp$ the scale associated with
$\zeta$, so that \begin{equation}
\lnp\sim\inv{\zeta}\sim\left\{ \begin{array}{ll}
16\pi^{2}\mnp & \mathrm{weakly}\mbox{\textrm{-}}\mathrm{coupled\, and\, decoupling\, heavy\, physics}\\
\mnp & \mathrm{strongly\, coupled\, heavy\, physics\,(NDA\, estimate)\,.}\end{array}\right.\label{eq:def.of.lnp}\end{equation}

\subsection{The Lagrangian in terms of mass eigenfields\label{sub:Mass-eigenstates}}

From $\lcal$ it is straightforward to obtain the neutrino and lepton
mass matrices and electroweak moments after SSB. Replacing $\phi\to\langle\phi\rangle=(v/\sqrt{2})(0,1)$
yields the following mass terms for the leptons\begin{equation}
\mathcal{L}_{m}=-\overline{e_{L}}M_{e}e_{R}-\overline{\nu_{L}^{\prime}}M_{D}\nu_{R}^{\prime}-\half\overline{\nu_{L}^{\prime c}}M_{L}\nu_{L}^{\prime}-\half\overline{\nu_{R}^{\prime c}}M_{R}\nu_{R}^{\prime}+\mathrm{h.c.}\label{eq:Lmass-prime}\end{equation}
\begin{equation}
M_{R}=M+\xi v^{2},\quad M_{L}=\chi v^{2},\quad M_{D}=Y_{\nu}\frac{v}{\sqrt{2}},\quad M_{e}=Y_{e}\frac{v}{\sqrt{2}};\end{equation}
it is worth noting that, up to possible coupling-constant factors,
$M_{D}\sim v$ while $M_{L}\sim v^{2}/\mnp$. Various situations obtain
depending on the hierarchy between $M_{R},\: M_{D}$ and $M_{L}$:
the standard (type I) see-saw scenario results from $M_{R}\gg M_{D}\gg M_{L}$;
types II and III see-saw are indistinguishable at the level of the
dimension 5 effective Lagrangian and correspond to $M_{L}\gg M_{D}^{2}/M_{R}$.
For these cases there is no conserved or approximately conserved fermion
number and the mass eigenstates are Majorana fermions. In contrast,
when $M_{D}\gg M_{R,L}$ there is an approximately conserved fermion
number and the mass eigenstates will be Dirac fermions up to small
admixtures (pseudo-Dirac case).

When $M_{R}\gg M_{D}\gg M_{L}$ the mass matrices can approximately
be diagonalized in blocks leading to two $3\times3$ Majorana mass
matrices \begin{eqnarray}
\mathrm{heavy}: &  & \mcal_{N}\approx M_{R}\,,\\
\mathrm{light}: &  & \mcal_{\nu}\approx M_{L}-M_{D}^{*}\inv{M_{R}^{\dagger}}M_{D}^{\dagger}\,.\end{eqnarray}
These matrices can subsequently be diagonalized by using the unitary
matrices $U_{N}$ and $U_{\nu}$, $M_{N}=U_{N}^{T}\mathcal{M}_{N}U_{N}$
and $M_{\nu}=U_{\nu}^{T}\mathcal{M}_{\nu}U_{\nu}$ with $M_{N}$ and
$M_{\nu}$ diagonal matrices with positive elements (in general one
can choose $\mathrm{\mathcal{M}}_{N}$ diagonal, in which case $U_{N}=1$).
Thus, the mass terms (\ref{eq:Lmass-prime}) can be rewritten in terms
of mass eigenfields as (without loss of generality we can also take
$M_{e}$ real and diagonal with positive elements)\[
\mathcal{L}_{m}=-\bar{e}M_{e}e-\half\bar{\nu}M_{\nu}\nu-\half\overline{N}M_{N}N\,,\]
and the $\nu_{L,R}^{\prime}$ have simple expressions in terms of
the light ($\nu$) and heavy ($N$) mass-eigenstate Majorana fields
($\nu=\nu^{c}$ and $N=N^{c}$) \begin{eqnarray}
\nu_{L}^{\prime} & = & P_{L}\left(U_{\nu}\nu+\varepsilon U_{N}N+\cdots\right);\label{eq:nuL2nuN}\\
\nu_{R}^{\prime} & = & P_{R}\left(U_{N}N-\varepsilon^{T}U_{\nu}\nu+\cdots\right);\label{eq:nuR2nuN}\end{eqnarray}
with $P_{L,R}=(1\mp\gamma_{5})/2$ the usual chirality projectors,
and \begin{equation}
\varepsilon\approx M_{D}M_{R}^{-1}\label{eq:heavy-light-mixing}\end{equation}
a $3\times3$ matrix characterizing the mixing between heavy and light
neutrinos. Note that barring cancellations in $\mathcal{M}_{\nu}$,
the elements of the mixing matrix $\varepsilon$ in eqs.~(\ref{eq:nuL2nuN}--\ref{eq:heavy-light-mixing})
obey generically ($m_{\nu}$ is a mass of the order of the light neutrino
masses and $m_{N}$ a mass of the order of the heavy neutrino masses)
\begin{equation}
|\varepsilon_{ij}|\lesssim\sqrt{\frac{m_{\nu}}{m_{N}}};\label{eq:def.of.mix}\end{equation}
 leading to a strong suppression of all mixing effects in most scenarios.

Substituting eq.~(\ref{eq:nuR2nuN}) in eq.~(\ref{eq:L5}) and using
the well know expression of $B_{\mu}$ in terms of the photon and
the $Z$ field, we obtain the relevant interactions in terms of the
mass eigenfields. For instance from the right-handed electroweak moment
interaction we obtain\begin{equation}
\mathcal{L}_{\zeta}=\left(\overline{N}U_{N}^{\dagger}-\overline{\nu}U_{\nu}^{\dagger}\varepsilon^{*}\right)\sigma^{\mu\nu}\left(\zeta P_{R}+\zeta^{\dagger}P_{L}\right)\,\left(U_{N}N-\varepsilon^{T}U_{\nu}\nu\right)\left(c_{W}F_{\mu\nu}-s_{W}Z_{\mu\nu}\right)\,,\label{eq:zeta-interactions}\end{equation}
where $F_{\mu\nu}$ and $Z_{\mu\nu}$ are the Abelian field strengths
of the photon and the $Z$-gauge boson respectively, and $c_{W}=\cos\theta_{W}$
, $s_{W}=\sin\theta_{W}$ with $\theta_{W}$ the weak mixing angle. 

We see that the $\nu_{R}^{\prime}$ electroweak moment operator generates
a variety of couplings when expressed in terms of mass eigenstates.
These vertices include a tensor coupling of the $Z$-boson and magnetic
moment couplings for both $N$ and $\nu$, as well as $N-\nu$ transition
moments. Note, however, that there is a wide range in the magnitude
of the couplings, in particular heavy-light couplings are suppressed
by $\varepsilon$ and light-light couplings are suppressed by $\varepsilon^{2}$. 

Similarly, if we substitute eqs.~(\ref{eq:nuL2nuN}-\ref{eq:nuR2nuN})
in the last term of eq.~\ref{eq:L5} and choose the unitary gauge,
we obtain, in addition to a contribution to the $N$ mass, a Higgs-heavy
neutrino interaction: \begin{equation}
\mathcal{L}_{\xi}=-vH\overline{N}\left(\xi P_{R}+\xi^{\dagger}P_{L}\right)N+\cdots\,,\label{eq:xi-interactions}\end{equation}
where we again took $U_{N}=1$ and the dots represent other interactions
generated by this operator: $HHNN$ vertices as well as $N-\nu$ and
$\nu-\nu$ interactions that are suppressed by the mixing $\varepsilon$;
these vertices are also generated by the neutrino Yukawa coupling
in $\mathcal{L}_{\nu_{R}}$ and are also suppressed.

Finally we should mention that when eq.~(\ref{eq:nuL2nuN}) is substituted
in the SM weak interaction terms $\overline{\nu_{L}^{\prime}}\gamma^{\mu}\nu_{L}^{\prime}Z_{\mu}$
and $\overline{e_{L}}\gamma^{\mu}\nu_{L}^{\prime}W_{\mu}$, one obtains
$N\mathrm{-}\nu\mathrm{-}Z$, $N\mathrm{-}e\mathrm{-}W$ couplings,
which, although suppressed by $\varepsilon$, are important for the
decays of the lightest of the heavy neutrinos%
\footnote{One also generates a $Z\mathrm{-}N\mathrm{-}N$ coupling suppressed
by $\varepsilon^{2}$.%
}.

\section{Collider effects\label{sec:collider}}

The new heavy particles responsible for the right-handed electroweak
moment must be charged under the electroweak group and are then expected
to have standard couplings to the photon and the $Z$ gauge bosons.
Since they have not been produced at LEP2 or Tevatron we can conclude
that $M_{NP}>100\,\mathrm{GeV}$. As discussed above, if the new physics
is perturbative, the associated effective scale in the coupling $\zeta$
is $1/\zeta=(4\pi)^{2}M_{NP}>15\,\mathrm{TeV}$, and its effects will
be suppressed. However, it is possible for the new interactions to
be generated in the strong coupling regime%
\footnote{This scenario is in many aspects similar to the case of excited neutral
fermions which has been largely considered in the literature~(for
limits from LEP1 and LEP2 see for instance \cite{Abreu:1996vd,Decamp:1990sq}
and \cite{Abbiendi:1999sa,Abreu:1998jw}, and for prospects at future
colliders see \cite{Belyaev:2004qq,Eboli:2001hi}), with the difference
that in our case we have right-handed neutrinos which do not have
standard weak interactions.%
}, in which case $\zeta$ can be much larger and may have interesting
effects at near-future colliders such as the LHC. It is then worth
studying the effects of the new interactions for this scenario; accordingly,
following the estimates in eq.~\eqref{eq:def.of.lnp}, we will take
$\zeta=1/\Lambda_{NP}$ and study the impact of the new interactions
at LEP, LHC and ILC. The results for the perturbative regime can be
recovered by taking $\Lambda_{NP}=(4\pi)^{2}M_{NP}$. 

As discussed previously, the mixing between light and heavy neutrinos
is $\varepsilon\sim\sqrt{m_{\nu}/m_{N}}$, so that all effects $\propto\varepsilon$
will be negligible unless $m_{N}$ is very small, but for light $m_{N},$
$m_{N}<10\,\mathrm{KeV}$, we have very stringent bounds on the coupling
from astrophysical considerations which will render the effects at
colliders negligible (see section~\ref{sec:astro}). Thus, in most
cases all mixing effects can be ignored. The main exception occurs
when studying the decays of the lightest $N$ which becomes stable
when $\varepsilon=0$.

\subsection{Decay rates and decay lengths}

Before discussing the impact of the new interactions in past and future
colliders, we would like to discuss briefly the dominant decay modes
of the new neutral fermions and their decay lengths for the relevant
experiments. Although in principle we could have three or more right-handed
neutrinos, for simplicity we will only consider the two lightest ones,
$N_{1}$ and $N_{2}$ (with $m_{1}<m_{2}$). The extension to more
heavy neutral fermions is straightforward. 

If the magnetic-moment-type interactions are strong enough to produce
the new particles, the dominant decay modes of the heaviest neutrino,
$N_{2}$, will be $N_{2}\rightarrow N_{1}\gamma$, and $N_{2}\rightarrow N_{1}Z$
if the $N_{2}$ is heavy enough%
\footnote{If $N_{2}$ and $N_{1}$ are almost perfectly degenerate these decays
will be suppressed. In that case decays to SM particles like $N_{2}\rightarrow\nu\gamma$,
$N_{2}\rightarrow eW$, $N_{2}\rightarrow\nu Z$ or $N_{2}\rightarrow\nu H$,
although suppressed by $\varepsilon,$ could be relevant. %
}. For relatively heavy $N_{2}$, $m_{2}>10\,\mathrm{GeV}$, the produced
photons will be hard and can be measured. The lifetime will be very
small and the decay length very short; for example, we find that for
$N_{2}$ produced at center of mass (CM) energies ranging from $100-1000\,\mathrm{GeV}$,
the decay lengths of the $N_{2}$ are well below $10^{-8}\:\mbox{m}$
unless $m_{2}\approx m_{1}$. 

In contrast, the lightest heavy neutrino, $N_{1}$, must decay into
SM particles. As discussed above, this means that $N_{1}$ decays
will always be suppressed by the mixing parameter $\varepsilon$ (which
we take as $\varepsilon=10^{-6}$ for our estimates), and the corresponding
decay lengths will be much longer.

\begin{figure}[h]
\includegraphics[width=0.8\columnwidth]{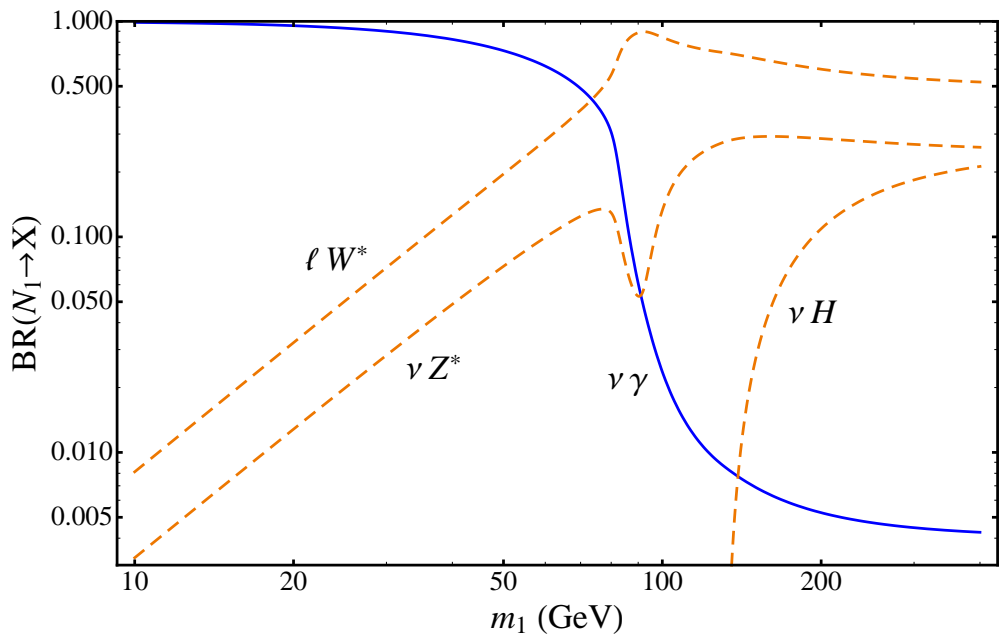}

\caption{Decay branching ratios of $N_{1}$. Solid for $N_{1}\rightarrow\nu\gamma$
and dashed for $N_{1}\rightarrow eW^{*}\rightarrow e+\mathrm{fermions}$,
$N_{1}\rightarrow\nu Z^{*}\rightarrow\nu+\mathrm{fermions}$ and $N_{1}\to\nu H$
(see text). We take $\varepsilon\sim10^{-6}$ , $\Lambda_{NP}=10\,\mathrm{TeV}$
and $m_{H}=130\,\mathrm{GeV}$. \label{fig:BRN1}}

\end{figure}

Since all the decay widths of the $N_{1}$ are proportional to $\varepsilon$,
the branching ratios will depend weakly on the heavy-light mixing
parameters; they will, however, be sensitive to the strength of the
new magnetic moment interaction. An example is presented in figure~\ref{fig:BRN1}
for $\Lambda_{NP}=10\,\mathrm{TeV}$: if $m_{1}<m_{W}$ the decay
is dominated by $N_{1}\rightarrow\nu\gamma$ although for larger masses
of the $N_{1}$ the tree-body decay $N_{1}\rightarrow eW^{*}\rightarrow e+\mathrm{fermions}$
could also be important.

\begin{figure}[h]
\includegraphics[width=0.8\columnwidth]{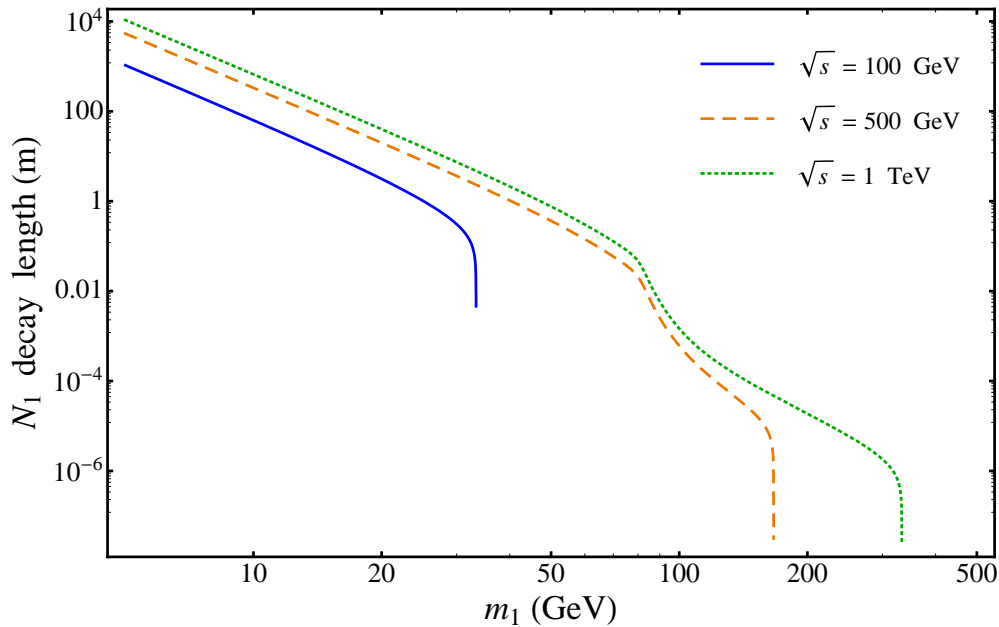}

\caption{$N_{1}$ decay lengths for a $N_{1}$ produced together with a $N_{2}$
at CM. We present results for CM energies of $\sqrt{s}=100\,\mathrm{GeV}$
(solid), $500\,\mathrm{GeV}$ (dashed), and $1\,\mathrm{TeV}$ (dotted);
we took $m_{2}=2m_{1}$, $\Lambda_{NP}=10\,\mathrm{TeV}$ and $\varepsilon=10^{-6}$
.\label{fig:N1decay-lengths}}

\end{figure}

For $m_{1}$ above $m_{W}$ the decays are dominated by the two body
decay $N_{1}\rightarrow\ell W$ and for masses above $m_{Z}$ the
decay $N_{1}\rightarrow\nu Z$ is also important%
\footnote{In this and several other points we disagree with the results presented
in~\cite{Kikuchi:2008ki}.%
}. If $m_{1}>m_{H}$, the $N_{1}$ can also decay into a real Higgs
boson (in the figure we have taken $m_{H}=130\,\mathrm{GeV}$), however
for these masses the Higgs boson width is very small, therefore virtual
production is suppressed and the branching ratio drops rapidly once
$m_{1}\lesssim m_{H}$. Notice that for $m_{1}\gg m_{H}$, the decay
widths $\Gamma(N_{1}\rightarrow\nu Z)$ and $\Gamma(N_{1}\rightarrow\nu H)$
are equal and half of $\Gamma(N_{1}\rightarrow eW)$, as required
by the equivalence theorem~\cite{Lee:1977eg,Cornwall:1974km} (see
also the discussion in appendix~\ref{sub:N1-decay-rates}). Notice
also that in figure~\ref{fig:BRN1} we have taken $\Lambda_{NP}=10\,\mathrm{TeV}$
and the decay width $\Gamma\left(N_{1}\rightarrow\nu\gamma\right)$
is suppressed by $1/\Lambda_{NP}^{2}$ while the decays to weak gauge
bosons are not. Thus, for relatively small $\Lambda_{NP}$, $\Lambda_{NP}\sim1\,\mathrm{TeV}$,
the decay $N_{1}\rightarrow\nu\gamma$ could also be relevant even
above the threshold of production of weak gauge bosons.

In figure~\ref{fig:N1decay-lengths} we present an estimate of the
$N_{1}$ decay lengths as a function of its mass. We assume that the
$N_{1}$ is produced through the new electroweak moment interaction
together with a $N_{2}$ (for instance $e^{+}e^{-}\rightarrow N_{1}N_{2}$)
at CM, and subsequently decays into the allowed channels, $N_{1}\to\nu V$
($V=\gamma,W,Z)$. Decay lengths are presented as a function of the
$N_{1}$ mass for different values of the CM energy for $m_{2}=2m_{1}$,
$\Lambda_{NP}=10\,\mathrm{TeV}$ and $\varepsilon=10^{-6}$. We observe
that the decay lengths of the $N_{1}$ will be very small for masses
above $100\,\mathrm{GeV}$. However, for masses below $100\,\mathrm{GeV}$
the decay lengths could range from a few millimeters to a few kilometers,
depending on the $N_{1}$ and the $N_{2}$ masses, the heavy-light
mixing, the electroweak coupling and the kinematical configuration
of the experiment. In particular there is an intermediate range of
masses for which the $N_{1}$ could be identified through the presence
of a displaced photon vertex~\cite{Terwort:2008ii,Zalewski:2007up,Prieur:2005xv}.

\subsection{Heavy neutrinos in $e^{+}e^{-}$ colliders}

As mentioned previously, if $N_{1}$ and $N_{2}$ are sufficiently
light, the fact that these particles were not observed at LEP1~\cite{Abreu:1996pa,Adriani:1992pq,Akrawy:1990zq,Dittmar:1989yg}
and LEP2~\cite{Abdallah:2003np,Achard:2003tx,Abbiendi:1998yu} places
strong bounds on their couplings. The most conservative bound is obtained
by assuming that both $N_{1}$ and $N_{2}$ escape undetected. This
is likely for a relatively light $N_{1}$ because it can only decay
through heavy-light mixing and, as discussed above, the corresponding
decay length could be very large. The $N_{2}$, however, will decay
into $N_{1}$ and $\gamma$, with the energetic photon providing a
potentially clear signature. In that case stronger bounds can be set
but those bounds will depend on the details of the spectrum%
\footnote{For instance, if the $N_{1}$ and $N_{2}$ are almost degenerate the
photon will be too soft to provide a viable signal.%
}. Instead of providing an exhaustive description of all possible scenarios
we will limit ourselves to the interesting case of the bounds that
can be derived from the LEP data when it is assumed that the $Z$
decays invisibly into $N_{1},N_{2}$; then, at the end of this section,
we will comment on the bounds that could be derived from visible $N_{2}$
decays.

The decay width $\Gamma\left(Z\rightarrow N_{1}N_{2}\right)$ is given
in appendix~\ref{sec:Decay-rates} and it is proportional to $|\zeta_{12}|^{2}$.
Assuming that only the standard decays $Z\rightarrow\nu_{\ell}\bar{\nu}_{\ell}\ \ (\ell=e,\mu,\tau)$
and $Z\rightarrow N_{1}N_{2}$ contribute to the invisible width of
the $Z$-boson, $\Gamma_{inv}$, we can obtain a bound on $|\zeta_{12}|$.
Using the experimental values~\cite{Amsler:2008zzb} we have

\begin{equation}
\Gamma_{inv}=3\Gamma_{\bar{\nu}\nu}^{SM}+\Gamma(Z\rightarrow N_{1}N_{2})=499.0\pm1.4\,\mathrm{MeV}\ .\label{invisible}\end{equation}
Using also the charged lepton $Z$ boson width, $\Gamma_{\bar{\ell}\ell}=83.984\pm0.086\,\mathrm{MeV}$
and the ratio of the neutrino and charged leptons partial widths calculated
within the SM, $\Gamma_{\bar{\nu}\nu}^{SM}/\Gamma_{\bar{\ell}\ell}^{SM}=1.991\pm0.001$,
we find \begin{equation}
\Gamma(Z\rightarrow N_{1}N_{2})=\Gamma_{inv}-3\left(\frac{\Gamma_{\bar{\nu}\nu}}{\Gamma_{\bar{\ell}\ell}}\right)^{SM}\Gamma_{\bar{\ell}\ell}\simeq-2.6\pm1.5\:\mathrm{MeV}\,.\label{dginv}\end{equation}
Since $\Gamma(Z\rightarrow N_{1}N_{2})$ is positive and the mean
value is negative, we use the Feldman \& Cousins prescription~\cite{Feldman:1997qc}
to estimate the $90\%$ CL bound\[
\Gamma(Z\rightarrow N_{1}N_{2})<0.48\times1.5\:\mathrm{MeV=0.72\:}\mathrm{MeV\qquad90\%}\:\mathrm{CL\,,}\]
which in our case implies that\begin{equation}
\Lambda_{NP}=\frac{1}{|\zeta_{12}|}>7\sqrt{f_{Z}(m_{Z},m_{1},m_{2})}\:\mathrm{TeV}\,,\label{eq:LEPBound}\end{equation}
where $f_{Z}(m_{Z},m_{1},m_{2})$ is a phase space factor given in
the appendix~\ref{sec:Decay-rates} normalized in such a way that
$f_{Z}(m_{Z},0,0)=1$. For example, $\Lambda_{NP}>1.9$ TeV if $m_{1}=m_{2}=35$~GeV.

\begin{figure}
\includegraphics[width=0.8\columnwidth]{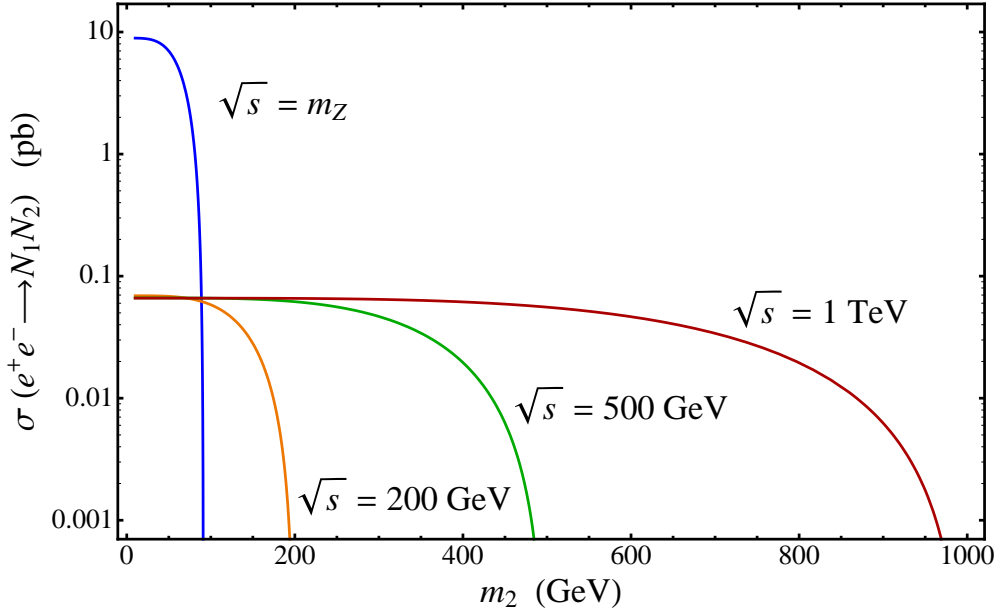}

\caption{$e^{+}e^{-}\rightarrow N_{1}N_{2}$ as a function of the heavy neutrino
mass, $m_{2}$, for different center of mass energies. We took $m_{1}=0$,
$\Lambda_{NP}=10\,\mathrm{TeV}$\label{fig:eeN1N2}}

\end{figure}

If the right-handed neutrino electroweak moment is large enough to
allow significant production of $N_{1},N_{2}$ pairs at LEP energies,
the dominant decay of $N_{2}$ will also be $N_{2}\rightarrow N_{1}\gamma$,
unless the mass of the $N_{1}$ is very close to the $N_{2}$ mass.
Then, the resulting photons could be detected and separated from the
background if $E_{\gamma}>10\,\mathrm{GeV}$. In fact, searches for
this type of processes (some searches for excited neutrinos also fall
in this class of processes) have been conducted at LEP1~\cite{Abreu:1996pa,Abreu:1996vd,Adriani:1992pq,Akrawy:1990zq}
and at LEP2~\cite{Abdallah:2003np,Achard:2003tx,Abbiendi:1998yu}.
If the mass of the heavy neutrino is below $\sim90\,\mathrm{GeV}$
one typically obtains upper bounds on the production branching ratio
$BR(Z\rightarrow N_{1}N_{2})$ of the order of $2\times10^{-6}$--$8\times10^{-6}$
(see for instance \cite{Abreu:1996vd,Abreu:1996pa}) depending on
the masses of $N_{1}$ and $N_{2}$ (these results also assume that
$BR(N_{2}\rightarrow N_{1}\gamma)=1$ and that $m_{2}>5\,\mathrm{GeV}$).
Using these data one can set much stronger bounds. For instance if
$m_{1}=0$ and $m_{2}$ is relatively light, $10\,\mathrm{GeV}<m_{2}<m_{Z}$,
we can use the conservative limit $BR(Z\rightarrow N_{1}N_{2})<8\times10^{-6}$
and obtain $\Lambda_{NP}=1/|\zeta_{12}|>40\:\mathrm{TeV}$. Data from
LEP2 can also be used to place limits~~\cite{Abdallah:2003np,Achard:2003tx,Abbiendi:1998yu}
on the couplings for masses up to $200\,\mathrm{GeV}$. For typical
values of $m_{1,2}$ one can set upper bounds on the production cross
section of the order of $0.1\,\mathrm{pb}$ (for $\sqrt{s}=207\,\mathrm{GeV}$)
which translate into bounds on $1/|\zeta_{12}|$ of the order of a
few TeV. LEP bounds based on visible $N_{2}$ decays depend more strongly
on the $N_{1}$ and $N_{2}$ masses (for instance, they are completely
lost lost if $m_{2}-m_{1}\lesssim10\,\mathrm{GeV}$) but they could
be important if some signal of this type is seen at the LHC.

In figure~\ref{fig:eeN1N2} we give the cross section for $e^{+}e^{-}\rightarrow N_{1}N_{2}$
as a function of $m_{2}$ (for illustration we took $m_{1}=0$ and
$\Lambda_{NP}=10\,\mathrm{TeV}$) for the center of mass energies
of LEP1 and LEP2 (we plotted values for $\sqrt{s}=200\,\mathrm{GeV}$).
We also included results for $\sqrt{s}=500\,\mathrm{GeV}$ and $\sqrt{s}=1000\,\mathrm{TeV}$
in view of the proposals for future $e^{+}e^{-}$ colliders as the
International Linear Collider (ILC). We see that, except for collisions
at the $Z$ peak, which are enhanced by about two orders of magnitude,
or close to the threshold of production, which are suppressed by phase
space, cross sections are quite independent on the CM energy and are
of the order of $0.1\,\mathrm{pb}$ for $\Lambda_{NP}=10\,\mathrm{TeV}.$

\subsection{Neutral heavy lepton production at the LHC}

\begin{figure}[h]
\includegraphics[width=0.8\columnwidth]{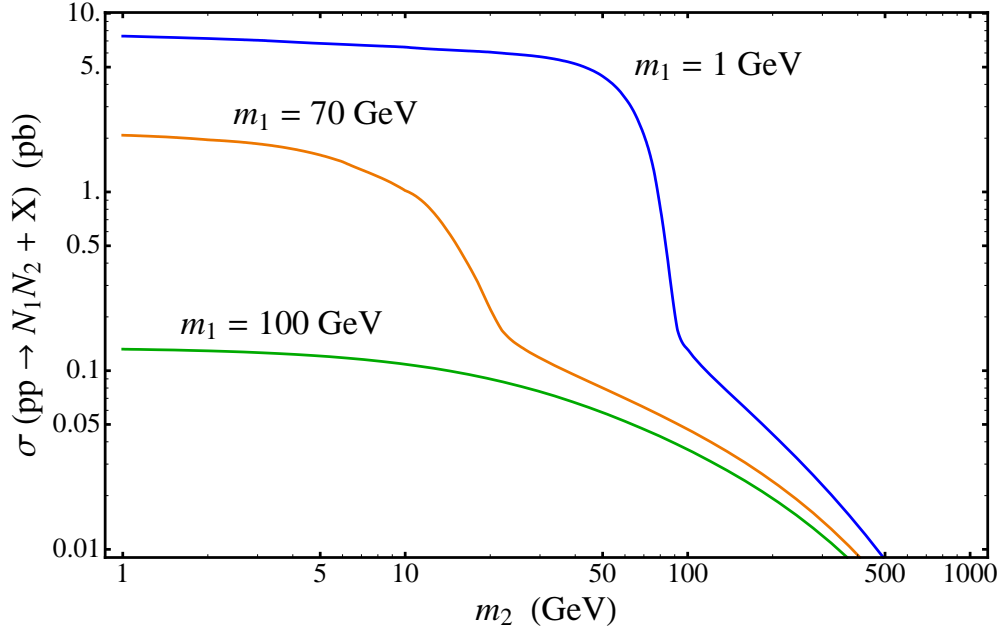}

\caption{$pp\rightarrow N_{1}N_{2}+X$ cross section at the LHC ($\sqrt{s}=14\,\mathrm{TeV})$
as a function of the mass of $N_{2}$. We took $\Lambda_{NP}=10\,\mathrm{TeV}$
and drew three curves for few representative masses of the $N_{1}$.\label{fig:ppN1N2}}

\end{figure}

The right-handed electroweak moment can help to produce the heavy
neutrinos at hadron colliders. In particular, heavy neutrinos will
be produced at the LHC through the Drell-Yan process. The differential
cross section for proton-proton collisions can be computed in terms
of the the partonic cross sections (for a very clear review see for
instance~\cite{Campbell:2006wx})

\[
d\sigma(pp\rightarrow N_{1}N_{2}+X)=\sum_{q}\int_{0}^{1}dx_{1}\int_{0}^{1}dx_{2}\left(f_{q}(x_{1},\hat{s})f_{\bar{q}}(x_{2},\hat{s})+(q\leftrightarrow\bar{q})\right)d\hat{\sigma}(q\bar{q}\rightarrow N_{1}N_{2},\hat{s})\,,\]
where $\hat{s}=x_{1}x_{2}s$ is the partonic center of mass invariant
square mass, $\hat{\sigma}$ is the partonic cross section and $f_{q}(x_{1},\hat{s})$,
$f_{\bar{q}}(x_{2},\hat{s})$ are the parton distribution functions
for the proton. Taking the partonic cross sections given in appendix~\ref{sec:Decay-rates}
and performing the convolution over the parton distribution functions
we find the total cross section as a function of the heavy neutrino
masses%
\footnote{We have used the CTEQ6M parton distribution sets~\cite{Pumplin:2002vw}.
One could also include next-to-leading-order corrections by multiplying
by a $K$-factor which typically would change cross sections by $10-20\%$.
Results have been checked against the CompHEP program~\cite{Boos:2004kh,Pukhov:1999gg}.%
}.

\begin{figure}[h]
\includegraphics[width=0.8\columnwidth]{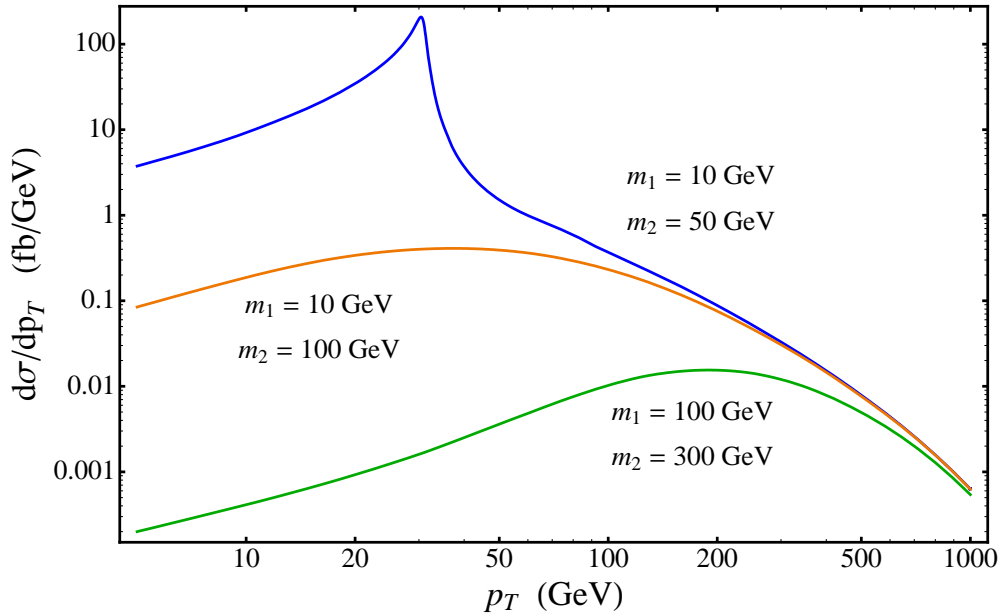}

\caption{Transverse momentum distribution of the process $pp\rightarrow N_{1}N_{2}+X$
for different sets of heavy neutrino masses.\label{fig:Differential-cross-sectionLHC}}

\end{figure}

The cross section depends on the masses and the coupling $\zeta_{12}=1/\Lambda_{NP}$.
In figure~\ref{fig:ppN1N2} we represent the total cross section
for $\Lambda_{NP}=10\,\mathrm{TeV}$ as a function of the $N_{2}$
mass for $\sqrt{s}=14\,\mathrm{TeV}.$ We give results for three representative
values of $m_{1}$. We see that cross sections above $100\:\mathrm{fb}$
are easily obtained but only for $m_{1}+m_{2}\lesssim m_{Z}$, where
LEP bounds apply. For larger masses the cross section decreases very
fast.

In figure~\ref{fig:Differential-cross-sectionLHC} we present the
differential cross section for the process $p\, p\rightarrow N_{1}\, N_{2}+X$
(with respect to the transverse momentum) for different sets of neutral
heavy lepton masses. For $m_{1}+m_{2}<m_{Z}$ we see clearly the peak
of the $Z$ gauge boson.

\subsection{Higgs decays into heavy neutrinos}

In this paper we are mainly interested in the effects of a possible
magnetic moment of right-handed neutrinos. However, as discussed before,
among the three possible dimension five operators there is one which
gives a correction to the right-handed neutrino Majorana mass. Moreover,
it also gives new Higgs boson couplings which could be relevant for
Higgs boson searches at the LHC/ILC. In particular, it could induce
new additional decays of the Higgs into right-handed neutrinos which
could be dominant in some region of parameters, particularly if the
Higgs mass is in the range $m_{H}\sim100-160\,\mathrm{GeV}$ and if
the right-handed neutrinos are light enough to be produced in Higgs
decays. Let us discuss briefly the possible effects of this operator. 

\begin{figure}[h]
\includegraphics[width=0.8\columnwidth]{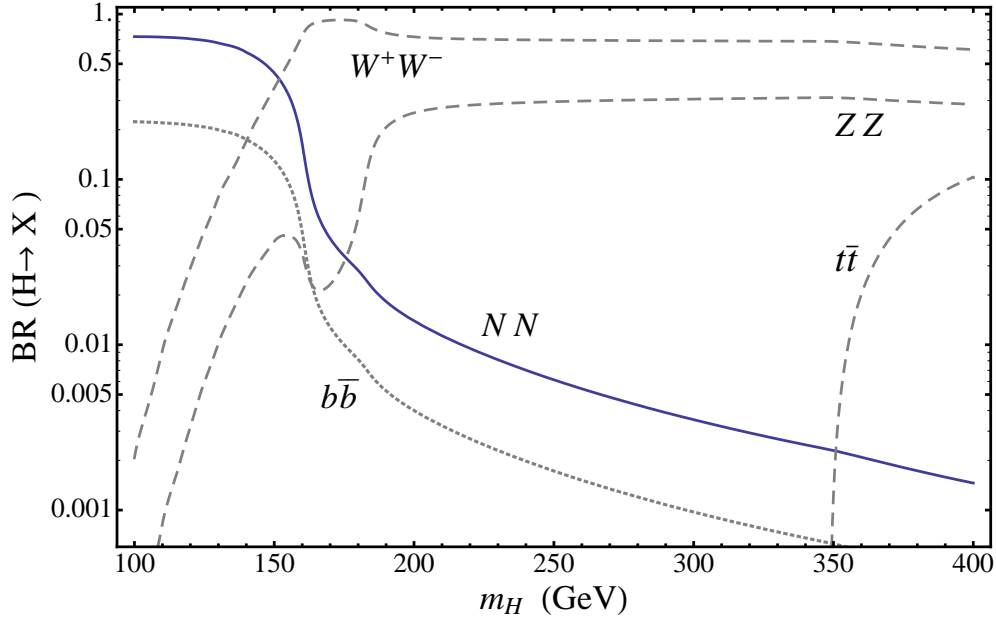} 

\caption{Estimated branching ratios for Higgs decays with the new-physics scale
at $1/\xi=10$~TeV. Heavy neutrino masses have been neglected.\label{fig:HiggsBR10}}

\end{figure}

In subsection~\ref{sub:Mass-eigenstates} we derived the relevant
interactions induced by the new operators. In particular the Higgs
boson interaction with heavy neutrinos is given in eq.~\eqref{eq:xi-interactions},
where couplings $H$-$\nu$-$N$ and $H$-$\nu$-$\nu$, which are
suppressed, have been neglected.

From eq.~\eqref{eq:xi-interactions} we compute the decay width of
the Higgs boson into two heavy neutrinos which is given in appendix~\ref{sec:Decay-rates}.
Then, we can compare with the SM decay rates of the Higgs boson. In
figure~\ref{fig:HiggsBR10} we represent the decay branching ratios
into the different channels for the new physics scale given by%
\footnote{Notice that this interaction can be generated at tree level; therefore,
up to possible small couplings, $1/\xi$ can be identified directly
with the masses of the new physics particles in the perturbative regime. %
} $M_{NP\xi}=1/\xi=10\,\mathrm{TeV}$. For simplicity we neglected
heavy neutrino masses. For heavier neutrinos there are some phase
space suppression factors given in appendix~\ref{sec:Decay-rates}.
We see that if $m_{H}$ lies below the $WW$ threshold, right-handed
neutrinos dominate Higgs decays (if kinematically allowed). In fact,
for low enough $M_{NP\xi}$, these decays could be significant even
when the $WW$ and $ZZ$ channels are open. This also means that the
branching ratios to other interesting channels in this region, as
for instance $H\rightarrow2\gamma$, are suppressed and could make
its detection more difficult. However, the effect of this new interaction
is not necessarily so bad since the produced $N's$ have to decay.
If the magnetic-moment interaction of right-handed neutrinos is also
present the heaviest neutrinos can decay into lighter ones and photons,
and those photons could be detected. Moreover, the lightest of the
heavy neutrinos will decay into light neutrinos and photons. As discussed
in section~\ref{sec:leff}, this is suppressed by the mixing heavy-light,
therefore the $N_{1}$ could be rather long-lived and produce non-pointing
photons which could be detected. If the magnetic moment interaction
is not present, the heavy neutrinos will have three-body decays ($N_{1}\rightarrow W^{*}\nu$
or $N_{1}\rightarrow Z^{*}\nu$) suppressed by the heavy-light mixing%
\footnote{Recall that the interesting mass range is $m_{H}<160\,\mathrm{GeV}$
and such a light Higgs boson cannot decay into real $W$'s or $Z$'s.
However, if one of the heavy neutrinos is light enough, then the heavier
one could still decay into real $W$'s and $Z$'s and produce interesting
signals. %
}.

\section{Astrophysical and cosmological considerations\label{sec:astro}}

In this section we consider several astrophysical and cosmological
systems and processes that may be affected by the presence of a magnetic
coupling of the neutrinos. Neither the calculations nor the list are
intended to be exhaustive; we will instead focus on some of the most
interesting effects.

\subsection{Astrophysical effects}

Among the various astrophysical processes that are affected by neutrino
magnetic couplings the cooling of red giant stars plays a prominent
role because it provides a very tight bound on the magnitude of the
magnetic moments -- provided the masses of the neutrinos involved
are sufficiently small. This limit is based on the observation that
in a plasma photons acquire a temperature-dependent mass (and are
then referred to as plasmons); any electromagnetic neutrino coupling
will then open a decay channel for the plasmon into a neutrino pair,
unless kinematically forbidden. If produced, the neutrinos leave the
star, resulting in an additional cooling mechanism that is very sensitive
to the size of the magnetic moment~\cite{Castellani:1993hs,Catelan:1996-461,Haft:1993jt,Raffelt:1989xu,Raffelt:1990pj,Raffelt:1992pi,Heger:2008er};
this can be used to impose stringent upper limit on this moment.

The electroweak moment couplings of mass eigenstates derived from
$\mathcal{L}_{5}$ is given in eq.~\eqref{eq:zeta-interactions}.
In particular the electromagnetic coupling of heavy neutrino eigenstates
is (we already took $U_{N}=1$). \begin{equation}
\mathcal{L}_{EM}=c_{W}\overline{N}\sigma^{\mu\nu}\left(\zeta P_{R}+\zeta^{\dagger}P_{L}\right)N\, F_{\mu\nu}\,.\label{eq:Bnu}\end{equation}
In a nonrelativistic nondegenerate plasma the emissivity of neutrinos
is dominated by transverse plasmons~\cite{Raffelt199605}, which
have an effective mass equal to the plasma frequency $\omega_{P}$.
A calculation shows that the decay width of these plasmons into two
neutrino species, labeled by $i$ and $j$ and satisfying $m_{i}+m_{j}<\omega_{P}$,
is \begin{equation}
\Gamma(\mathrm{plasmon}\rightarrow N_{i}N_{j})=\frac{2c_{W}^{2}\left|\zeta_{ij}\right|^{2}}{3\pi}\frac{\omega_{P}^{4}}{\omega}f_{Z}(\omega_{P},m_{i},m_{j})\,,\end{equation}
where $\omega$ is the plasmon energy in plasma rest frame, and $f_{Z}$
has been defined in eq.~(\ref{eq:fz}). The total decay rate is then
\[
\Gamma(\mathrm{plasmon}\to NN)=\frac{\mu_{{\rm \textrm{eff}}}^{2}}{24\pi}\frac{m_{P}^{4}}{\omega}\,,\]
\begin{equation}
\mu_{{\rm \textrm{eff}}}^{2}=16c_{{\rm W}}^{2}\sum_{\mathrm{all}}\left|\zeta_{ij}\right|^{2}f_{Z}(\omega_{P},m_{i},m_{j})\,,\label{eq:mu-plasmon}\end{equation}
and the sum runs over all allowed channels, $i>j$ such that $m_{i}+m_{j}<\omega_{P}$.
The observational limits from red giant stars cooling then imply~\cite{Raffelt199605}
\begin{equation}
\mu_{{\rm \textrm{eff}}}<3\times10^{-12}\mu_{B}\,,\end{equation}
where $\mu_{B}$ is the Bohr magneton. This translates into a bound
on the couplings $\zeta_{ij}$ \emph{provided} the sum of the associated
neutrino masses lies below $\omega_{P}$, for example, for $\zeta_{ij}$
real, \begin{equation}
|\zeta_{ij}|<8.5\times10^{-13}\mu_{B};\quad m_{i,j}\ll\omega_{P}\simeq8.6\,\mathrm{KeV\,.}\end{equation}
This then gives $\Lambda_{NP}\gtrsim4\times10^{6}\,\mathrm{TeV}$;
this bound is degraded somewhat when the neutrino masses are comparable
to $\omega_{P}$.

It is clear from eq.~\eqref{eq:zeta-interactions} that the photon
(plasmon) can also decay into $N$-$\nu$ and $\nu$-$\nu$. However,
the relevant couplings for these processes are suppressed by $\varepsilon$
and $\varepsilon^{2}$, respectively, which are small numbers (for
instance, if $m_{\nu}\sim0.1\,\mathrm{eV}$ and $m_{N}\sim1\,\mathrm{keV}$,
$\varepsilon\sim0.01$, see (\ref{eq:def.of.mix})). Therefore, this
mixing can only affect plasmon decays for extremely light $N,$ $m_{N}\sim m_{\nu}$;
in this case all neutrino masses can be neglected compared to the
plasma frequency $\omega_{P}\sim10\,\mathrm{keV}$, and since photons
only couple to right-handed neutrinos, our result still applies (taking
$m_{i}=m_{j}=0$). Alternatively, if $m_{N}>\omega_{P}\sim10\,\mathrm{keV}$
the heavy neutrinos cannot be produced in plasmon decay and the only
bound comes from $\mathrm{plasmon}\rightarrow\nu\nu$; however, the
amplitude for this process is suppressed by $\varepsilon^{2}$ which
is very small if $m_{N}\gg\omega_{P}\sim10\,\mathrm{keV},$ so that
the bounds derived from this process are weak (if we take $\varepsilon^{2}\sim m_{\nu}/m_{N}$
we roughly expect $\Lambda_{NP}\gtrsim(m_{\nu}/m_{N})\times4\times10^{6}\,\mathrm{TeV}\sim40\,\mathrm{TeV}$
for $m_{\nu}=0.1\,\mathrm{eV}$ and $m_{N}=10\,\mathrm{keV}$ and
drops below a TeV already for $m_{N}>0.4\,\mathrm{MeV}$).

The same type of reasoning can be applied to other astrophysical objects.
This might be of interest because the corresponding plasma frequency
$\omega_{P}$ will be larger in denser objects, so that the corresponding
limits will apply to heavier neutrino states; unfortunately the limits
themselves are much poorer. As an example, we consider the case of
a neutron star whose plasma frequency in the crust is $\omega_{P}\sim1\,\mathrm{MeV}$.
This could allow us to extend the magnetic moment bounds to higher
neutrino masses; however, the much weaker limit, $\mu_{{\rm eff}}<5\times10^{-7}\mu_{B}$~\cite{Iwamoto:1994zd}
implies $\Lambda_{NP}\gtrsim23\,\mathrm{TeV}$ when $m_{i,j}\lesssim1\,\mathrm{MeV}$
which is not competitive with bounds derived below from $\gamma+\nu\to N$
in supernovas, which also apply in this range of masses. Limits derived
for plasmon decays from solar and supernova data are also not competitive~\cite{Raffelt199605,Raffelt:1999gv}.

The neutrino electromagnetic coupling would also affect other interesting
processes. For example, it generates a new supernova cooling mechanism
through $\gamma+\nu\to N$ (when kinematically allowed), with the
$N$ escaping. Limits on this {}``anomalous'' cooling~\cite{Raffelt199605}
imply that the effective magnetic moment then must lie below $3\times10^{-12}\mu_{B}$
provided the heavy neutrino mass lies below $\sim30\,\mathrm{MeV}$
(which is of the order of the maximum neutrino energy in the supernova
core). The coupling for this process is suppressed $\sim\zeta\varepsilon\sim\left(\sqrt{m_{\nu}/m_{N}}\right)/\Lambda_{NP}$,
so we find $\Lambda_{NP}\gtrsim4\times10^{6}\times\sqrt{m_{\nu}/m_{N}}\,\mathrm{TeV}$.
Taking, for example, $m_{\nu}\sim0.1\,\mathrm{eV}$ we obtain $\Lambda_{NP}>1.5\times10^{4}\,\mathrm{TeV}$
for $m_{N}=10\,\mathrm{keV}$ and $\Lambda_{NP}>390\,\mathrm{TeV}$
for $m_{N}=10\,\mathrm{MeV}$. These limits are interesting in the
region $10\,\mathrm{keV<}m_{N}<30\,\mathrm{MeV}$, where red giant
bounds do not apply. 

It is also worth noting that if the $N$ mass is $m_{N}\sim1\,\mathrm{keV}$,
these particles may contribute to the dark matter content of the universe~\cite{Dodelson:1993je,Shi:1998km}
(but see also \cite{Seljak:2006qw,Viel:2006kd,Lesgourgues:2006nd,Melchiorri:2008gq}).
However, although the bounds on the right-handed neutrino magnetic
moment coming from red giants apply, they could still have important
effects in the analysis and further study is necessary.

\subsection{CP asymmetries}

The electroweak moments involving only the $\nu_{R}^{\prime}$ are
also of interest because they generate lepton number violation and
may contribute to the baryon asymmetry of the universe~\cite{Fukugita:1986hr}.
Though providing a complete description of these effects lies beyond
the scope of the present paper we will provide a simplified discussion
of the issues involved.

\begin{figure}[ht]
\begin{centering}
\includegraphics[width=0.9\columnwidth]{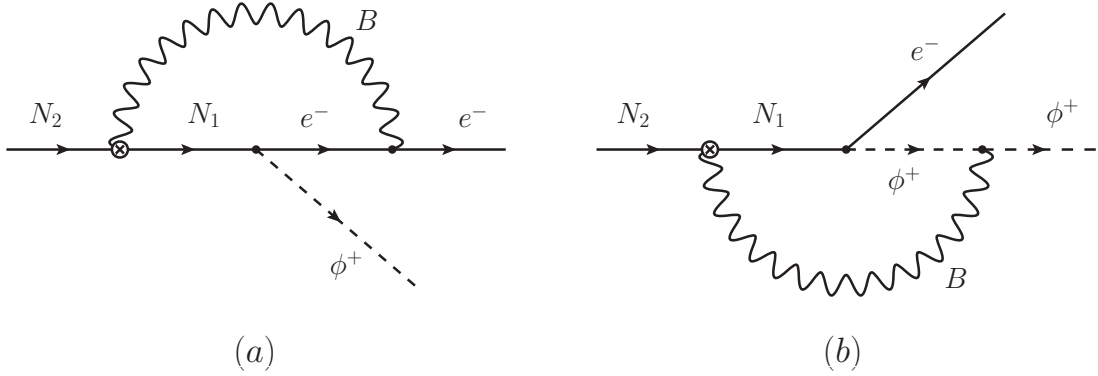} 
\par\end{centering}

\caption{1-loop graphs involving electroweak moments contributing to $L$-violating
heavy-neutrino decays}

\label{fig:l.viol} 
\end{figure}

In the presence of the electroweak moments the relevant lepton-number-violating
decays remain the standard%
\footnote{For a review of leptogenesis together with references to the original
literature see, for example, ref.~\cite{Davidson:2008bu}. For new
mechanism of leptogenesis involving neutrino magnetic moments see~\cite{Bell:2008fm}
and for leptogenesis using composite neutrinos see~\cite{Grossman:2008xb}.%
} $N\to e^{\pm}\phi^{\mp}$ (here $e^{\pm}$ denotes a charged lepton
and $\phi^{\mp}$ the charged scalar components of the Higgs doublet)
which receive a contribution from this dimension five operator. The
new graphs, however, necessarily involve a virtual heavy neutrino
$N^{\prime}$ (see fig. \ref{fig:l.viol}) and will generate a lepton
asymmetry only if $N^{\prime}$ is lighter than $N$. Because of this,
this type of contributions may be relevant only when the lightest
of the heavy neutrino states are degenerate or almost degenerate (for
a recent review talk on these scenarios see~\cite{Pilaftsis:2009pk}).

The calculation of the contributions of the Majorana electroweak moments
to the lepton-number-violating decay width of the $N$ is straightforward.
We will assume that $m_{N}\gg v$ so that we can neglect electroweak
symmetry breaking and assume that all gauge bosons, leptons and scalars
are massless except the heavy neutrino which has a Majorana mass term.
Also, for simplicity, we neglect Yukawa couplings for charged leptons.
The relevant pieces of the Lagrangian are discussed in section~\ref{sec:leff},
in particular in eqs.~(\ref{eq:LagrangianSM}--\ref{eq:L5}) and
(\ref{eq:xi-interactions}), \begin{eqnarray}
\mathcal{L}_{N} & = & \frac{i}{2}\overline{N}\sla{\partial}N-\half\overline{N}M_{N}N-\overline{\ell}Y_{\nu}P_{R}N\tilde{\phi}-\tilde{\phi}^{\dagger}\overline{N}Y_{\nu}^{\dagger}P_{L}\ell+\overline{N}\sigma^{\mu\nu}(\zeta P_{R}+\zeta^{\dagger}P_{L})N\, B_{\mu\nu}\,,\label{eq:LagrangianN}\end{eqnarray}
where $N$ are Majorana fields and $M_{N}$ is their mass matrix which,
without loss of generality, can be taken diagonal. Since we ignore
the charged lepton Yukawa couplings we can rotate the doublet fields
$\ell$ so that $Y_{\nu}$ is Hermitian; there are no other possible
field redefinitions so $\zeta$ is, in general, antisymmetric and
complex. For $n$ generations both $Y_{\nu}$ and $\zeta$ contain
$n(n-1)/2$ phases; in particular, for $n=3$ we will have a total
of $6$ phases. But even for $n=2$ we have two phases since both
$Y_{12}$ and $\zeta_{12}$ can be complex. This is important because
$CP$--violating observables should depend on those couplings; it
also means that we can make our estimates in a model with just $2$
generations, as we will do for simplicity.

Assuming 2 generations with $N_{2}$ the heavier of the right-handed
neutrinos, we consider the lepton-number-violating decays $N_{2}\to e^{-}\phi^{+}$
and $N_{2}\to e^{+}\phi^{-}$. At tree level the amplitudes are simply
\begin{eqnarray}
\mathcal{A}_{0}(N_{2} & \to e^{-}\phi^{+})= & Y_{e2}\bar{u}(p_{e})P_{R}u(p_{2})\,,\\
\mathcal{A}_{0}(N_{2} & \to e^{+}\phi^{-})= & Y_{e2}^{*}\bar{v}(p_{2})P_{L}v(p_{e})=-Y_{e2}^{*}\bar{u}(p_{e})P_{L}u(p_{2})\,,\end{eqnarray}
 where we used $v(p)=u^{c}(p)$.

The one-loop corrections to these processes induced by the electroweak
moment coupling $\zeta$ are given in figure~\ref{fig:l.viol}. Notice
that if the external particle is $N_{2}$ then the antisymmetry of
$\zeta$ dictates that only $N_{1}$ can run in the loop. Thus, if
$m_{2}>m_{1}$ we expect (finite) imaginary contributions from these
graphs. A straightforward but tedious calculation confirms this expectation.
Explicitly we find the following CP-violating asymmetry in $N_{2}$
decays to be\[
\epsilon_{\notcp}\equiv\frac{\Gamma(N_{2}\rightarrow e^{-}\phi^{+})-\Gamma(N_{2}\rightarrow e^{+}\phi^{-})}{\Gamma(N_{2}\rightarrow e^{-}\phi^{+})+\Gamma(N_{2}\rightarrow e^{+}\phi^{-})}\]
\begin{equation}
=-\frac{g^{\prime}}{2\pi}(m_{2}^{2}-m_{1}^{2})\frac{m_{1}}{m_{2}^{3}}\Imag{\frac{Y_{e2}Y_{e1}^{*}}{|Y_{e2}|^{2}}\left(\zeta_{12}^{*}m_{2}+\zeta_{12}m_{1}\right)}\,.\label{eq:CPasymmetry}\end{equation}
For $m_{1}\ll m_{2}$\[
\epsilon_{\notcp}=-\frac{g^{\prime}}{2\pi}m_{1}\Imag{\frac{Y_{e2}Y_{e1}^{*}}{|Y_{e2}|^{2}}\zeta_{12}^{*}}\sim-\frac{g^{\prime}}{2\pi}\frac{m_{1}}{\Lambda_{NP}}\Imag{\frac{Y_{e2}Y_{e1}^{*}}{|Y_{e2}|^{2}}e^{-i\delta_{12}}}\,,\]
where $\delta_{12}$ is the phase of $\zeta_{12}.$

We see that the Majorana electroweak moments do generate additional
contributions to CP violating asymmetries in heavy neutrino decays.
These, however, are relevant only for the decay of the heavier neutrinos
and so could be relevant for leptogenesis only when $m_{1}$ and $m_{2}$
are relatively close~\cite{Covi:1996wh,Flanz:1996fb,Pilaftsis:1997jf,Pilaftsis:2003gt}.
In this limit the amplitude is proportional to $(m_{2}^{2}-m_{1}^{2})$;
despite this suppression the possible relevance of these interactions
requires a careful comparison of all contributions, and this lies
beyond the scope of the present investigation.

\section{Summary of bounds, prospects and conclusions\label{sec:conclusions}}

As can be seen from the previous sections, the dimension 5 operators
involving right-handed neutrinos open up observable effects in several
scenarios of interest. The electroweak moment operator (first term
in eq.~\eqref{eq:L5}) provides the richest phenomenology, but contributions
coming from the $\left(\phi^{\dagger}\phi\right)\overline{\nu_{R}^{\prime c}}\xi\nurp$
operator (last term in eq.~\eqref{eq:L5}) can affect Higgs boson
decays. After spontaneous symmetry breaking, this operator gives rise
to several interaction vertices involving right-handed neutrinos and
the Higgs boson, the strongest being a simple $H\: N_{i}N_{j}$ term,
which provides new decay channels of the Higgs to $N$'s (if such
a process is kinematically allowed). These decays could dramatically
change the Higgs decay branching ratios (see figure~\ref{fig:HiggsBR10}),
especially in the region $100\,\mathrm{GeV}<m_{H}<160\,\mathcal{\mathrm{GeV}}$
where the gauge boson channels are still closed. The new decays could
result in an invisible Higgs, if the heavy neutrinos cannot be detected,
or in new, enhanced detection channels if the right-handed neutrinos
can be seen through their own decay channels, for instance $N_{2}\rightarrow N_{1}\gamma$,
or $N_{1}\rightarrow\nu\gamma$ and $N_{1}\rightarrow eW$ with a
displaced vertex.

\begin{figure}[h]
\includegraphics[width=0.8\columnwidth]{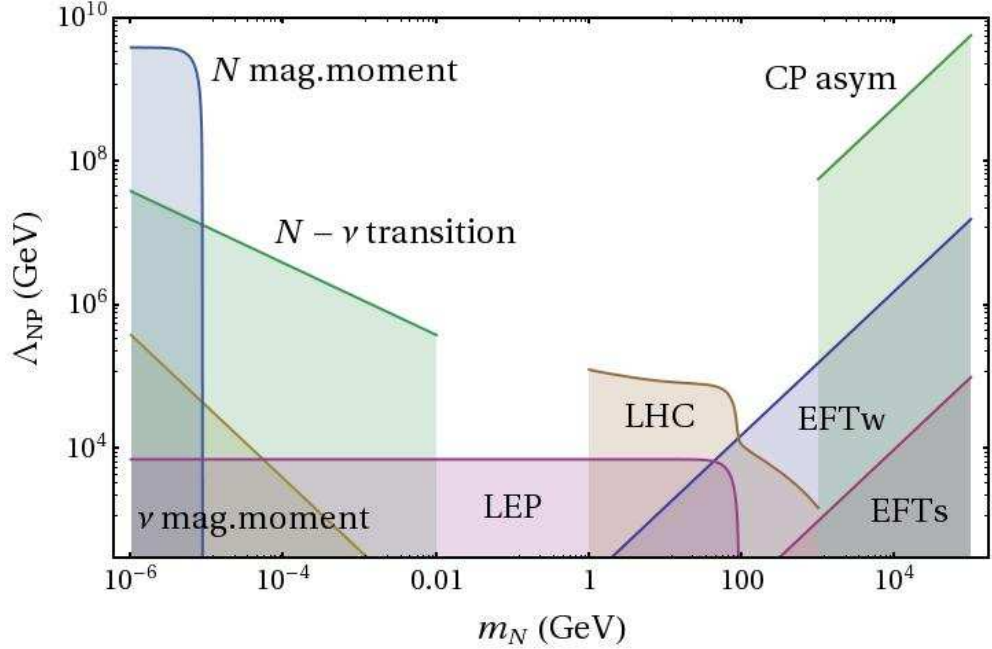}

\caption{Summary of bounds and prospects. The shaded areas labeled $\nu,N$
mag. moment, $N-\nu$ transition and LEP denote regions excluded by
the corresponding observables; the areas marked EFTw and EFTs correspond
to the regions where the EFT parametrization is inconsistent (for
the weak- and strong-coupling regimes, respectively). Finally, shaded
areas marked CP asym. and LHC denote the range of parameters where
the dimension 5 electroweak moment might affect the corresponding
observables. See text for details. \label{fig:Summary-of-bounds}}

\end{figure}

As for the electroweak moment operator, figure~\ref{fig:Summary-of-bounds}
summarizes present bounds on the model parameters as well as two regions
of potential interest, namely: the region relevant for the LHC and
the region that can provide a relatively large CP asymmetry. 

When expanded in terms of mass eigenstates the unique electroweak
moment operator generates $N-N$, $N-\nu$ and $\nu-\nu$ magnetic
moments, and $N-N$, $N-\nu$ and $\nu-\nu$ tensor couplings to the
$Z$-bosons, eq.~\eqref{eq:zeta-interactions}, giving rise to a
very rich phenomenology which depends basically on three parameters:
the coupling, $\zeta=1/\Lambda_{NP}$, the heavy-light mixing $\varepsilon$,
and the masses of the $N$. For our estimates in figure~\ref{fig:Summary-of-bounds}
we take $m_{N}=m_{2}$ and $\varepsilon\sim\sqrt{m_{\nu}/m_{N}}$
with $m_{\nu}=0.1\,\mathrm{eV}$, and neglect $m_{1}$. Then we represent
the regions in the $\Lambda_{NP}-m_{N}$ plane forbidden by the red
giant bound on the $N$ and $\nu$ magnetic moments, by the supernova
bound on the transition magnetic moment $N-\nu$ and by the LEP bound
from the {}``invisible'' $Z$-boson decay width.

To test the new interactions at the LHC one should produce first the
heavy neutrinos and then one should detect them. The analysis of the
detection is complicated and depends on the details of the spectrum
and the capabilities of the detectors, but at least one should produce
them with reasonable rates. Thus we require that the cross section
of $pp\rightarrow N_{1}N_{2}X$ is at least $100\,\mathrm{fb}$. 

The new interactions we have introduced contain new sources of CP
non-conservation which can modify the standard leptogenesis scenarios.
In particular we have found that the electroweak moment operator gives
additional contributions to the CP asymmetry in $N_{2}\rightarrow e^{-}\phi^{+}$
decays. These could be relevant in leptogenesis if $\epsilon_{\notcp}\sim(g^{\prime}/2\pi)m_{N}/\Lambda_{NP}>10^{-6}$
and $m_{N}>1\,\mathrm{TeV}$, a region that has also been represented
in figure~\ref{fig:Summary-of-bounds}.

Note that for the regions marked LHC and CP asymmetries the shadowed
area represents the region \emph{of interest}, in contrast to the
previous ones, for which the shadowed area represents the \emph{excluded}
region. 

Finally, the effective theory we use cannot be applied for all energies
and all masses. Thus, to give graphically an idea of the regions where
the EFT cannot be applied, we represent the regions with $m_{N}>\,\Lambda_{NP}$,
for the strong-coupling regime (EFTs) and $m_{N}>(4\pi)^{2}\Lambda_{NP}$,
for the weak-coupling regime (EFTw). 

From figure~\ref{fig:Summary-of-bounds} we can draw the following
conclusions:
\begin{lyxlist}{00.00.0000}
\item [{i)}] There are very tight bounds coming from red giants cooling
for $m_{N}\lesssim10\,\mathrm{keV}$, so strong as to require $\Lambda_{NP}>4\times10^{9}\,\mathrm{GeV}$;
in this scenario, obviously, any effect of the electroweak moment
coupling would be totally negligible in any present or planned collider
experiment.
\item [{ii)}] For $10\,\mathrm{keV}\lesssim m_{N}\lesssim10\,\mathrm{MeV}$
supernova cooling produced by the magnetic moment transitions $\gamma\,\nu\rightarrow N$,
provides very strong bounds. These bounds, however, depend on the
assumptions made on the heavy-light mixing parameter, $\varepsilon$.
For this mass range the magnetic moment limits from red giants are
derived from plasmon decay into a $\nu$ pair, which is proportional
to $\varepsilon^{2}$ and yields less restrictive constraints.
\item [{iii)}] For $m_{N}\lesssim m_{Z}$, the invisible $Z$ decays impose
$\Lambda_{NP}\gtrsim7\times10^{3}\,\mathrm{GeV}$, depending on the
details of the heavy neutrino spectrum%
\footnote{Most likely, searches for hard photons in the Galaxy X-ray background
could impose tighter bounds for this mass range, but the precise constraint
will depend on the details of the neutrino spectrum; a thorough examination
of this issue lies outside the scope of the present paper.%
}. 
\item [{iv)}] For $m_{N}\sim1\mathrm{-200\,}\mathrm{GeV}$ and roughly
$7\,\mathrm{TeV<}\Lambda_{NP}<100\,\mathrm{TeV}$, heavy neutrinos
could be produced at the LHC with cross sections above $100\,\mathrm{fb}$.
The heaviest two of them would decay rapidly to hard photons which
could be detected. The lightest one is quite long-lived and, in part
of the parameter space, would produce non-pointing photons which could
be detected.
\end{lyxlist}
Above we have expressed our conclusions in terms of $\Lambda_{NP}=1/\zeta$.
Since our operator is a magnetic moment-type operator, this scale
can only be interpreted as the mass of new particles in a non-perturbative
context. If it is generated by perturbative physics it arises at one
loop and one expects $\zeta\sim1/((4\pi)^{2}M_{NP})$, where $M_{NP}$
are the masses of the particles running in the loop and coupling constants
have been set to one. Thus, in this case, all the constraints discussed
above still apply to $M_{NP}=\Lambda_{NP}/(4\pi)^{2}$. Then, if the
new physics is weakly coupled, the interesting range for collider
physics, $\Lambda_{NP}\sim10\mathrm{-}100\,\mathrm{TeV}$ translates
into $M_{NP}\sim100-1000\,\mathrm{GeV}$. For such low masses the
effective theory cannot be applied at LHC energies and one should
use the complete theory that gives rise to right-handed neutrino electroweak
moments. Those models should contain new particles carrying weak charges
with masses $\sim100\mathrm{-}1000\,\mathrm{GeV}$ which should be
produced in the LHC via, for instance, the Drell-Yan process. 

As for the future work around this effective theory, much work still
remains to be done, especially concerning astrophysical and cosmological
scenarios:
\begin{lyxlist}{00.00.0000}
\item [{a)}] The magnetic coupling may have effects in the early universe
because it can potentially alter the equilibrium conditions of the
$N$ and their decoupling temperature. 
\item [{b)}] Heavy neutrinos with masses $\sim1\,\mathrm{keV}$ could be
a good dark matter candidate. The right-handed neutrino magnetic moments
could change significantly the analysis of this possibility. 
\item [{c)}] One should evaluate carefully the effects of the Majorana
magnetic couplings on non-thermal leptogenesis. 
\item [{d)}] For sufficiently large $\zeta$, this same coupling might
lead to the trapping of the right-handed neutrinos in the supernova
core. \end{lyxlist}
\begin{acknowledgments}
This work has been supported in part by the Ministry of Science and
Innovation (MICINN) Spain, under the grant number FPA2008-03373, by
the European Union within the Marie Curie Research \& Training Networks,
MRTN-CT-2006-035482 (FLAVIAnet), and by the U.S. Department of Energy
grant No.~DE-FG03-94ER40837. A.A. is supported, in part, by a FPU
fellowship from the MICINN. A.A and A.S. thank Sacha Davidson and
Jordi Vidal interesting discussions in the early stages of this work,
Vicente Vento for his advice on some of the calculations presented
here and Carmen Garcia for her help with the experimental data on
neutral heavy leptons; J.W. gratefully acknowledges Ernest Ma's help
and comments.
\end{acknowledgments}
\appendix

\section{Model calculation\label{sec:mod.calc}}

The simplest model that can generate a magnetic moment for the right-handed
neutrinos consists of adding to the standard model a vector-like fermion
$E$ and a scalar $\omega$, both isosinglets of hypercharge $y$,
with interactions \begin{equation}
\lcal_{\mathrm{int}}=\sum_{i}\lambda'_{i}\overline{\nu_{iR}^{c}}E\omega^{*}+\lambda_{i}\bar{E}\nu_{iR}\omega+\mathrm{h.c.}\end{equation}
 where $i$ is a family index; we take $\lambda_{i},\lambda'_{j}$
real. In this model the effective magnetic coupling is given by \begin{equation}
\zeta_{ij}=\frac{g'y(\lambda_{i}\lambda'_{j}-\lambda'_{i}\lambda_{j})}{64\pi^{2}\mnp};\qquad\mnp=m_{E}\frac{2(1-r)^{2}}{1-r+r\ln r},\qquad r=\left(\frac{m_{\omega}}{m_{E}}\right)^{2}\,.\end{equation}
 This choice of $\mnp$ is, of course, somewhat arbitrary, since experiment
only measures $\zeta$. We have chosen it so that $\mnp=m_{E}$ when
$m_{\omega}=m_{E}$).

\section{Decay rates and cross sections\label{sec:Decay-rates}}

Here we present the relevant formulas for decay rates and cross sections
used in the text. Before we introduce some notation useful to to simplify
the presentation of the formulas. First $s_{W}=\sin\theta_{W}$, $c_{W}=\cos\theta_{W}$
are the sine and cosine of the weak mixing angle. As usual we denote
by $q_{f}$ the charge of fermion $f$ and its vector, $v_{f}=t_{3}(f)\left(1-4|q_{f}|s_{W}^{2}\right)$,
and axial couplings, $a_{f}=t_{3}(f)$, with $t_{3}(f)=+1/2\,(-1/2)$
for up-type (down-type) fermions. We will write the new couplings
as $\zeta_{ij}=|\zeta_{ij}|e^{i\delta_{ij}}$. We will also define
as usual the Källen's Lambda function \begin{equation}
\lambda(a,b,c)=a^{2}+b^{2}+c^{2}-2ab-2ac-2bc\,.\label{eq:def.of.lambda}\end{equation}

\subsection{$Z\rightarrow N_{i}N_{j}$ }

The decay width of the $Z$ boson into heavy neutrinos is 

\begin{equation}
\Gamma(Z\rightarrow N_{i}N_{j})=\frac{2\abs{\zeta_{ij}}^{2}}{3\pi}s_{W}^{2}m_{Z}^{3}f_{Z}(m_{Z},m_{i},m_{j})\,,\end{equation}
where $f_{Z}(m_{Z},m_{i},m_{j})$ is a kinematical factor $f_{Z}(m_{Z},0,0)=1$\begin{equation}
f_{Z}(m_{Z},m_{i},m_{j})=\frac{\sqrt{\lambda(m_{Z}^{2},m_{i}^{2},m_{j}^{2})}}{m_{Z}^{6}}\left[m_{Z}^{2}\left(m_{Z}^{2}+m_{i}^{2}+m_{j}^{2}-6m_{i}m_{j}\cos2\delta_{ij}\right)-2\left(m_{i}^{2}-m_{j}^{2}\right)^{2}\right]\,.\label{eq:fz}\end{equation}

\subsection{$N_{2}$ decay rates}

If the new interaction is strong enough the dominant decays of the
heaviest neutral lepton proceed through the new interaction. The decay
rates are

\begin{equation}
\Gamma(N_{2}\rightarrow N_{1}\gamma)=\frac{2}{\pi}c_{W}^{2}\abs{\zeta_{12}}^{2}m_{2}^{3}\left(1-m_{1}^{2}/m_{2}^{2}\right)^{3}\,,\end{equation}
\begin{equation}
\Gamma(N_{2}\rightarrow N_{1}Z)=\frac{2}{\pi}s_{W}^{2}\abs{\zeta_{12}}^{2}m_{2}^{3}f_{2}(m_{Z},m_{1},m_{2})\,,\end{equation}
with \begin{equation}
f_{2}(m_{Z},m_{1},m_{2})=-\frac{m_{Z}^{6}}{2m_{2}^{6}}f_{Z}(m_{Z},m_{1},m_{2})\:,\qquad f_{2}(0,0,m_{2})=1\,.\end{equation}

\subsection{$N_{1}$ decay rates \label{sub:N1-decay-rates}}

The lightest of the heavy neutrinos, $N_{1}$, can decay only due
to mixing with the SM sector. If $m_{1}>m_{Z}$ the dominant decays
proceed through SM interactions induced by the mixing of heavy-light
neutrinos. \begin{equation}
\Gamma\left(N_{1}\rightarrow\ell_{\beta}^{-}W^{+}\right)=\frac{1}{16}\left|\varepsilon_{W}^{\beta1}\right|^{2}\,\frac{\alpha m_{1}^{3}}{s_{W}^{2}m_{W}^{2}}\,\left(1-\frac{m_{W}^{2}}{m_{1}^{2}}\right)^{2}\left(1+2\frac{m_{W}^{2}}{m_{1}^{2}}\right)\,.\end{equation}
 Here $\beta$ is a flavour index and $\varepsilon_{W}$ characterizes
the mixing of heavy-light neutrinos in $W$ boson couplings, which
is order $\sqrt{m_{\nu}/m_{N}}$. 

For $N_{1}\rightarrow\nu\, Z$ decays we obtain

\begin{equation}
\Gamma\left(N_{1}\rightarrow\nu_{\beta}Z\right)=\frac{1}{16}\left|\varepsilon_{Z}^{\beta1}\right|^{2}\,\frac{\alpha m_{1}^{3}}{s_{W}^{2}\, c_{W}^{2}m_{Z}^{2}}\,\left(1-\frac{m_{Z}^{2}}{m_{1}^{2}}\right)^{2}\left(1+2\frac{m_{Z}^{2}}{m_{1}^{2}}\right)\,,\end{equation}
 with $\varepsilon_{Z}$ is defined as above but for $Z$ boson couplings.
Notice that since $m_{W}=c_{W}m_{Z}$ the two decay widths are equal
up to phase space factors and differences in the mixing factors $\varepsilon_{Z}$
and $\varepsilon_{W}$. However, we have two decay channels into $W$'s,
$N_{1}\rightarrow e^{-}W^{+}$ and $N_{1}\rightarrow e^{+}W^{-}$,
and only one into $Z$'s (we already took into account that the $\nu_{\beta}$
are Majorana particles; should we treat them as Weyl particles, we
have two decay channels and the sum over them gives the same result). 

If $m_{1}>m_{H}$ the $N_{1}$ can also decay into Higgs bosons, $N_{1}\rightarrow\nu\, H$
with a decay width given by

\begin{equation}
\Gamma\left(N_{1}\rightarrow\,\nu_{\beta}\, H\right)=\frac{|Y_{\nu}^{\beta1}|^{2}m_{1}}{32\pi}\,\left(1-\frac{m_{H}^{2}}{m_{1}^{2}}\right)^{2}\,.\end{equation}
This looks quite different from $\Gamma\left(N_{1}\rightarrow\ell_{\beta}^{-}W^{+}\right)$
and $\Gamma\left(N_{1}\rightarrow\nu_{\beta}Z\right)$; however, we
can use that $\varepsilon\approx M_{D}M_{N}^{-1}$, $M_{D}=Y_{\nu}v/\sqrt{2}$
and $\alpha/(s_{W}^{2}m_{W}^{2})=1/(\pi v^{2})$ to rewrite $|\varepsilon|^{2}\alpha m_{1}^{3}/(s_{W}^{2}m_{W}^{2})\sim|Y_{\nu}|^{2}m_{1}/(2\pi v^{2})$
and see that, in the limit $m_{1}\gg m_{H},m_{W},m_{Z}$, the three
decay widths are identical. This is required by the equivalence theorem~\cite{Lee:1977eg,Cornwall:1974km}
which states that, in this limit, the calculation could have been
performed in the theory before spontaneous symmetry breaking; in that
theory, all the fields except the $N$ are massless, there is no heavy-light
mixing and the $N$'s decay into the doublet of leptons and the Higgs
scalar doublet through the standard model Yukawa couplings. However,
for moderate $m_{1}$, since $m_{H}>m_{Z}>m_{W}$, the phase space
factors are important; in particular $\Gamma(N_{1}\rightarrow\nu_{\beta}H)$
decreases rapidly when approaching the threshold of production. 

If $m_{1}<m_{W}$ the dominant decay is the decay into a light neutrino
and a photon. It requires the new interaction and light-heavy mixing.
\[
\Gamma\left(N_{1}\rightarrow\nu_{\beta}\gamma\right)=\frac{2}{\pi}\,\left|\varepsilon_{\gamma}^{\beta1}\right|^{2}\, c_{W}^{2}\, m_{1}^{3}\,,\]
 where $\varepsilon_{\gamma}$ is a parameter that characterizes the
strength of the $N_{1}$-$\nu_{\beta}$-$\gamma$ interaction and
it is of the order of $(1/\Lambda_{NP})\sqrt{m_{\nu}/m_{N}}$.

\subsection{$e^{+}e^{-}\rightarrow N_{1}N_{2}$ cross section}

By neglecting the heavy-light mixing, the LEP and ILC cross section
is given by

\begin{equation}
\sigma\left(e^{+}\, e^{-}\rightarrow N_{1}\, N_{2}\right)=\frac{2\alpha}{3}\left|\zeta_{12}\right|^{2}f_{Z}(\sqrt{s},m_{1},m_{2})\eta_{\ell}(s)\label{eq:fermion-cs}\end{equation}
with $f=e$,\begin{equation}
\eta_{f}(s)=4q_{f}^{2}c_{W}^{2}-4q_{f}v_{f}\mathrm{Re}\{\chi(s)\}+\frac{v_{f}^{2}+a_{f}^{2}}{c_{W}^{2}}\abs{\chi(s)}^{2}\,,\label{eq:etaf}\end{equation}
and\[
\chi(s)=\frac{s}{s-m_{Z}^{2}+im_{Z}\Gamma_{Z}}\,.\]

\subsection{Partonic cross sections for $pp\rightarrow N_{1}N_{2}X$}

To compute the $pp\rightarrow N_{1}N_{2}+X$ cross section we need
the different partonic cross sections $q\bar{q}\rightarrow N_{1}N_{2}$
which proceed through the new interaction and are dominated by $\gamma$
and $Z$ exchange.\[
\frac{d\hat{\sigma}}{d\Omega}\left(q\bar{q}\rightarrow N_{1}N_{2}\right)=\frac{\alpha}{6\pi}\left|\zeta_{12}\right|^{2}\eta_{q}(\hat{s})\frac{\sqrt{\lambda\left(\hat{s},m_{1}^{2},m_{2}^{2}\right)}}{\hat{s}^{3}}\]
\begin{equation}
\times\left[(m_{1}^{2}+m_{2}^{2})\,(\hat{s}+2\hat{t})-2\hat{t}\,(\hat{s}+\hat{t})-(m_{1}^{4}+m_{2}^{4})-2\hat{s}\, m_{1}m_{2}\cos2\delta_{12}\right]\,,\end{equation}
with $\hat{s}$ and $\hat{t}$ the Mandelstam variables for the partonic
collision in the center of mass frame of the quarks, and $\eta_{q}(\hat{s})$
is defined in eq.~\eqref{eq:etaf} with the quantum numbers appropriate
to the quarks. The total partonic cross section is obtained by integration
of the angular variables and leads to the result in eq.~\eqref{eq:fermion-cs}
with an additional factor $1/3$ due to color and with $q_{f}$,$a_{f}$,$v_{f}$
appropriate for $f=u,d$.

\subsection{Higgs boson decays into right-handed neutrinos $H\rightarrow N_{1}N_{2}$}

Above we have discussed only cross sections and decays induced by
the electroweak moment interaction or by standard model interactions
and heavy-light mixing. The last term in eq.~\ref{eq:L5} can also
have interesting consequences, in particular if the $N$'s are light
enough it can induce new decay modes for the Higgs boson. We found 

\begin{equation}
\Gamma\left(H\rightarrow N_{1}N_{2}\right)=\frac{v^{2}}{2\pi\, m_{H}^{3}}\,\abs{\xi_{12}}^{2}\sqrt{\lambda(m_{H}^{2},m_{1}^{2},m_{2}^{2})}\left[(m_{H}^{2}-m_{1}^{2}-m_{2}^{2})\,-2m_{1}m_{2}\cos2\delta_{12}^{\prime}\right]\,,\end{equation}
where $\xi_{ij}=\abs{\xi_{ij}}e^{i\delta_{ij}^{\prime}}$ and $v=\sqrt{2}\langle\phi^{(0)}\rangle$.

\bibliographystyle{h-physrev}
\bibliography{nureff}

\end{document}